\documentclass[%
 longbibliography,
 prapplied,%
 amsmath,amssymb,
 reprint,%
 superscriptaddress
,floatfix
]{revtex4-2}

\usepackage{graphicx}
\usepackage{siunitx}
\usepackage[utf8]{inputenc}
\usepackage{amsmath}
\usepackage{amsfonts}
\usepackage{amssymb}
\usepackage{natbib}
\usepackage{xcolor}
\usepackage{graphicx} 
\usepackage{braket}
\graphicspath{{Figs}}
\makeatletter
\newcommand{\manualsublabel}[3]{(#2)\def\@currentlabel{\ref{#3}(#2)}\label{#1}}
\DeclareUnicodeCharacter{FB01}{}
\makeatother

\begin{document}

\title{Exciton transport in a germanium quantum dot ladder}
\date{\today}

\author{T.-K. Hsiao}
\altaffiliation[Current address: ]{Department of Physics, National Tsing Hua University, Hsinchu 30013, Taiwan}
\email{These authors contributed equally to this work}
\affiliation{QuTech and Kavli Institute of Nanoscience, Delft University of Technology, 2600 GA Delft, The Netherlands}
\author{P. Cova Fariña}
\email{These authors contributed equally to this work}
\affiliation{QuTech and Kavli Institute of Nanoscience, Delft University of Technology, 2600 GA Delft, The Netherlands}
\author{S. D. Oosterhout}
\affiliation{QuTech and Kavli Institute of Nanoscience, Delft University of Technology, 2600 GA Delft, The Netherlands}
\affiliation{Netherlands Organisation for Applied Scientific Research (TNO), 2628 CK Delft, The Netherlands}
\author{D. Jirovec}
\affiliation{QuTech and Kavli Institute of Nanoscience, Delft University of Technology, 2600 GA Delft, The Netherlands}
\author{X. Zhang}
\affiliation{QuTech and Kavli Institute of Nanoscience, Delft University of Technology, 2600 GA Delft, The Netherlands}
\author{C. J. van Diepen}
\altaffiliation[Current address: ]{Niels Bohr Institute, University of Copenhagen, Copenhagen, Denmark}
\affiliation{QuTech and Kavli Institute of Nanoscience, Delft University of Technology, 2600 GA Delft, The Netherlands}
\author{W. I. L. Lawrie}
\altaffiliation[Current address: ]{Niels Bohr Institute, University of Copenhagen, Copenhagen, Denmark}
\affiliation{QuTech and Kavli Institute of Nanoscience, Delft University of Technology, 2600 GA Delft, The Netherlands}
\author{C.-A. Wang}
\affiliation{QuTech and Kavli Institute of Nanoscience, Delft University of Technology, 2600 GA Delft, The Netherlands}
\author{A. Sammak}
\affiliation{QuTech and Kavli Institute of Nanoscience, Delft University of Technology, 2600 GA Delft, The Netherlands}
\affiliation{Netherlands Organisation for Applied Scientific Research (TNO), 2628 CK Delft, The Netherlands}
\author{G. Scappucci}
\affiliation{QuTech and Kavli Institute of Nanoscience, Delft University of Technology, 2600 GA Delft, The Netherlands}
\author{M. Veldhorst}
\affiliation{QuTech and Kavli Institute of Nanoscience, Delft University of Technology, 2600 GA Delft, The Netherlands}
\author{E. Demler}
\affiliation{Institute for Theoretical Physics, Wolfgang Pauli Str. 27, ETH Zurich, 8093 Zurich, Switzerland}
\author{L. M. K. Vandersypen}
\email{L.M.K.Vandersypen@tudelft.nl}
\affiliation{QuTech and Kavli Institute of Nanoscience, Delft University of Technology, 2600 GA Delft, The Netherlands}

\begin{abstract}
Quantum systems with engineered Hamiltonians can be used as simulators of many-body physics problems to provide insights beyond the capabilities of classical computers. Semiconductor gate-defined quantum dot arrays have emerged as a versatile platform for quantum simulation of generalized Fermi-Hubbard physics, one of the richest playgrounds in condensed matter physics. In this work, we employ a germanium 4$\times$2 quantum dot array and show that the naturally occurring long-range Coulomb interaction can lead to exciton formation and transport. We tune the quantum dot ladder into two capacitively-coupled channels and exploit Coulomb drag to probe the binding of electrons and holes. Specifically, we shuttle an electron through one leg of the ladder and observe that a hole is dragged along in the second leg under the right conditions. This corresponds to a transition from single-electron transport in one leg to exciton transport along the ladder. Our work paves the way for the study of excitonic states of matter in quantum dot arrays.
\end{abstract}

\maketitle

\section{Introduction}
Analog quantum simulators with well-controlled interaction parameters can shed light on the physics of strongly-correlated many-body quantum systems~\cite{Cirac2012,Georgescu2014}. Electrostatically-defined semiconductor quantum dot arrays, owing to their in-situ tunability of electrochemical potentials and relevant energy scales which can far exceed the thermal energy, have become an attractive platform for simulating fermionic systems~\cite{Barthelemy2013,Manousakis2002,Byrnes2008,Yang2011}. Over the past few years, the techniques for control and probing of quantum dot simulators has progressed significantly. This platform and closely related donor arrays have been used as a small-scale simulator of Mott-Hubbard physics~\cite{Hensgens2017,Singha2011,Salfi2016}, Nagaoka ferromagnetism~\cite{Dehollain2020}, Heisenberg antiferromagnetic spin chains~\cite{VanDiepen2021}, resonating valence bond states~\cite{Wang2023}, and the Su–Schrieffer–Heeger model~\cite{Kiczynski2022}.

The charge carriers confined in quantum dot arrays exhibit an intrinsic long-range Coulomb interaction, which plays an important role in fundamental physics phenomena like exciton formation~\cite{Frenkel1931} and chemical bonding~\cite{French2010,Knorzer2022}, as well as in exotic phases such as Wigner crystals~\cite{Wigner1934,Vu2020}, excitonic insulators~\cite{Jerome1967} and exciton condensates~\cite{KOHN1970}. In contrast, simulating these phenomena is challenging for the highly successful quantum simulation platform based on ultra-cold atoms, where the inter-particle interaction is largely limited to on-site~\cite{Bloch2008,Arguello-Luengo2019}, or non-local dipole-dipole interaction~\cite{Chomaz2023,Browaeys2020,Bohn2017}.

One manifestation of the long-range Coulomb interaction in low-dimensional systems is Coulomb-drag. In a two-channel system, a current imposed by a voltage bias across one channel (the drive channel) leads to a current or voltage across a second channel (the drag channel)~\cite{Nandi2012}. Coulomb drag can take two forms. ``Positive" Coulomb drag occurs when an electron in the drive channel pushes electrons in the drag channel forward due to Coulomb-mediated momentum transfer~\cite{Gramila1991}. ``Negative" Coulomb drag can result from Wigner-crystal physics~\cite{Yamamoto2006} or from exciton formation~\cite{Narozhny2016, Nandi2012}, in which the motion of a charge carrier in the drive channel pulls along a charge carrier of opposite sign in the drag channel. The negative Coulomb drag effect from exciton formation has been observed in double quantum wells in the quantum hall regime~\cite{Kellogg2002,Tutuc2004,Nandi2012}, double quantum wires~\cite{Laroche2011} and 2D materials~\cite{Gorbachev2012,Li2017,Liu2017}. In these works, the negative Coulomb drag is interpreted as resulting from inter-channel exciton transport, which serves as a precursor for exciton condensation and excitonic insulator phases.

\begin{figure*} 
\centering    
\includegraphics[width=1\textwidth]{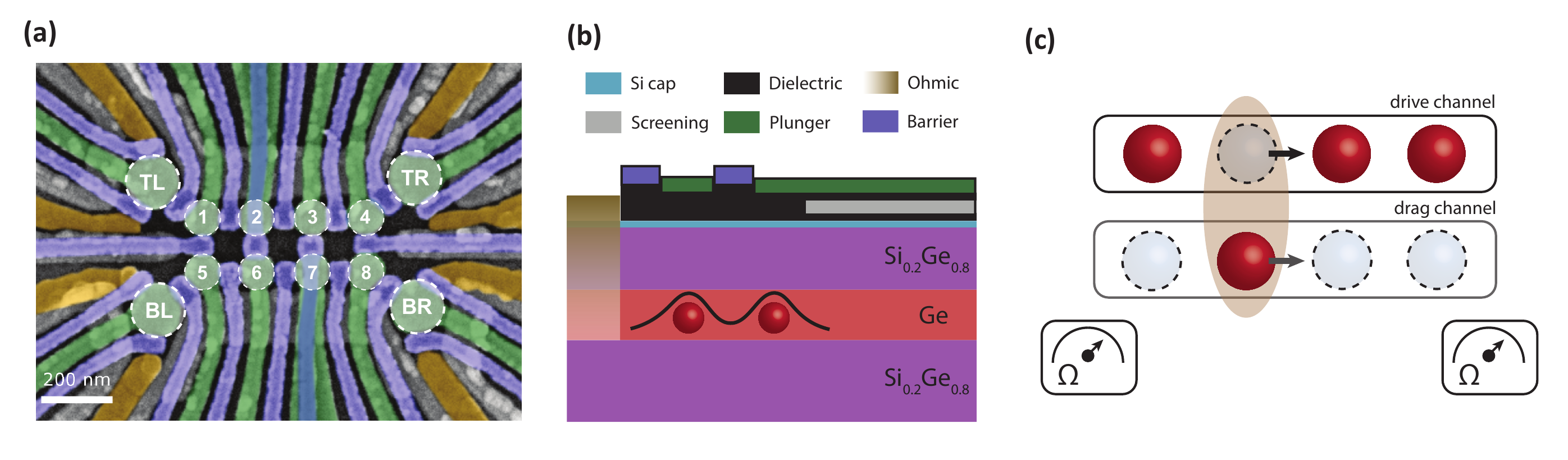}
\caption{(a) A false-color scanning electron microscope image of a device nominally identical to the one used in this work. The dashed white circles indicate the intended positions of the 4$\times$2 dot array and sensing dots. (b) Schematic cross-section of the gate stack and a germanium quantum well heterostructure. Holes are confined in the 55 nm-deep quantum well. Gate layers with different functions are drawn in colors shown in the legend. (c) Schematics illustrating the Coulomb drag of a hole in the bottom channel by the (imposed) motion of an electron (missing hole) in the top channel of a 4$\times$2 quantum dot array. The bound state of an electron and hole (an exciton) arises from the inter-channel Coulomb interaction. Two charge sensors, located at the bottom-left and bottom-right corners, are used to probe the charge configuration in the two array.}
\label{Device_fig}
\end{figure*}

Excitonic states can be described theoretically using a two-channel Hubbard model with $N\times 2$ sites~\cite{Pandey2021,Kaneko2012,vu2023excitonic}
\begin{equation}
\begin{aligned}
H =& -t\sum_{\langle i,j \rangle, \sigma} c^{\dagger}_{i\sigma}c_{j\sigma} + U\sum_{i}\frac{n_{i}(n_{i}-1)}{2}\\ &+ U^{\prime}\sum_{\langle i,j \rangle}n_{i}n_{j} + V\sum_{i\in \alpha,j\in \beta}n_{i}n_{j}\\ &+ V^{\prime}\sum_{i\in \alpha,j\in \beta}n_{i}n_{j}\\
\end{aligned}
\label{H_ex_eq}
\end{equation}
where $c_{i\sigma}$ denotes the annihilation operator of a spin-1/2 fermion with spin $\sigma$ $\in\{\uparrow$, $\downarrow$\} at site $i$ of a two-channel system where site $1$ to $N$ are located in channel $\alpha$ and site $(N+1)$ to $2N$ are part of channel $\beta$, and $\langle i,j \rangle$ sums over neighbouring sites in the same channel. The number operator is given by $n_{i}=c^{\dagger}_{i\uparrow}c_{i\uparrow}+c^{\dagger}_{i\downarrow}c_{i\downarrow}$,
$t$ is the tunnel coupling within the same channel, $U$ the on-site Coulomb interaction, $U^{\prime}$ is the nearest-neighbor Coulomb interaction within the same channel, $V$ is the nearest-neighbor inter-channel Coulomb interaction, and $V^{\prime}$ is the diagonal inter-channel Coulomb interaction. When the two channels are occupied by charge carriers of opposite sign, the inter-channel interactions are attractive. Note that we consider systems without hopping between the two channels and interaction terms beyond nearest-neighbor or diagonal sites are neglected. Furthermore, in Eq.~\ref{H_ex_eq} we assume homogeneous tunnel couplings and Coulomb interactions. To describe systems with inhomogeneous couplings, we will use $t_{ij}$ and $V_{ij}$ to denote the tunnel coupling and inter-channel Coulomb interaction between site $i$ and $j$. 

This model can describe the conduction band and valence band in a material, and also two capacitively-coupled channels. Earlier works have reported on arrays of metallic or superconducting tunnel junctions~\cite{Averin1991, Matters1997, Shimada2000}, and small quantum-dot arrays~\cite{Shinkai2009}. However, these systems lack the control knobs for individual interaction parameters and the probes for the quantum state at each site. In comparison, when a $N\times 2$ quantum dot ladder is tuned to host electrons in one channel and holes in the other channel, thanks to the advanced control and probing capabilities, it can be used as a versatile analog quantum simulator for excitonic physics.

Many years of work on quantum dot systems have led to steady scaling of linear arrays~\cite{Zajac2016,Philips2022,Ha2022}. Furthermore, several reports on two-dimensional quantum dot arrays have appeared using GaAs~\cite{Dehollain2020,Mortemousque2021}, silicon~\cite{Chanrion2020, Unseld2023} and germanium~\cite{Hendrickx2021} as the host material. Among the various host materials, germanium is particularly promising to scale to large arrays thanks to the low disorder and light effective mass~\cite{Lodari2019,Scappucci2021}. Even a $4 \times 4$ Ge quantum dot array has been realized~\cite{Borsoi2022}, albeit with shared-controlled electrochemical potentials and tunnel couplings.

In this work we use a 4$\times$2 Ge quantum dot ladder as an excitonic simulator, doubling the size of fully controlled Ge quantum dot arrays~\cite{Hendrickx2021}. We activate hopping along the legs of the ladder but suppress hopping between the legs. In this way, two capacitively-coupled channels of quantum dots are formed. The charge carriers in this platform are holes arising from the valence band. A missing hole on top of a singly-occupied background of holes effectively defines an electron.  We control the electrochemical potentials of the array such that the top channel hosts an electron and the bottom channel can host a hole. To explore the formation of excitons, we use real-time charge sensing to study under what conditions the imposed motion of an electron through the top channel drags along a hole in the bottom channel through the long-range Coulomb-interaction.

\section{Device and experimental approach}

\begin{figure*}
\centering    
\includegraphics[width=0.6\textwidth]{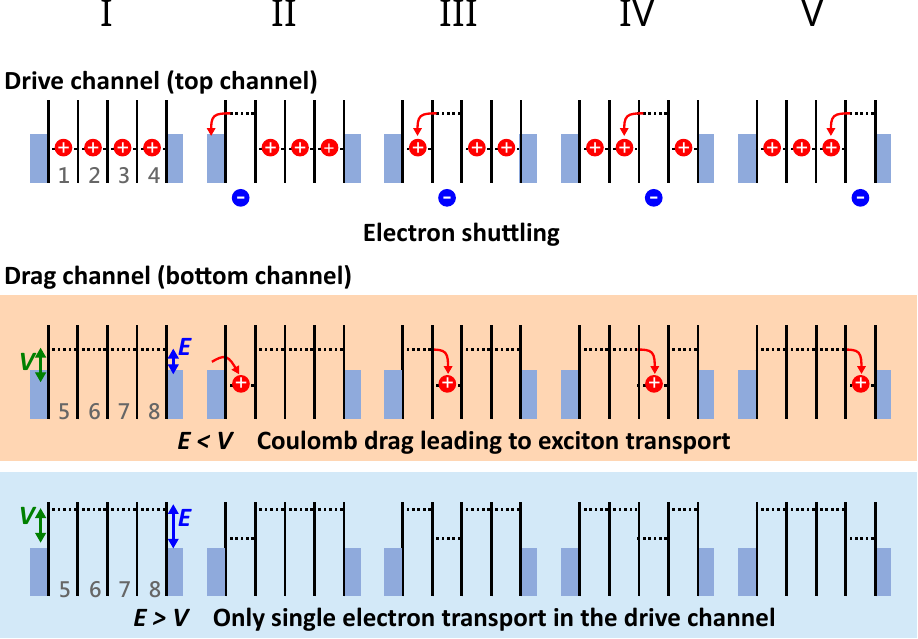}
\caption{Schematics illustrating the experimental scheme for probing exciton formation through Coulomb drag. The first row shows the ladder diagrams of the drive channel. The vertical axis is an energy axis, the horizontal axis is space. Vertical black lines indicate tunnel barriers, the blue shaded region represents the reservoir, which is filled up to the Fermi level. Phase I: Each dot is filled with one hole and the dot potentials are aligned. Phase II: The leftmost hole is pushed out by raising $\epsilon_1$, which can be viewed as loading an electron in dot 1. Phase III: $\epsilon_1$ is lowered and $\epsilon_2$ is raised. The electron moves to dot 2. Phase IV: $\epsilon_2$ is lowered and $\epsilon_3$ is raised. The electron moves to dot 3. Phase V: $\epsilon_3$ is lowered and $\epsilon_4$ is raised. The electron moves to dot 4. The second and the third rows compare the ladder diagrams of the drag channel with and without Coulomb drag effect, respectively. The second row is the exciton transport regime ($E<V_{ij}$), in which the presence of a drive-channel electron lowers the dot potentials in the drag channel sufficiently for a hole to be loaded in the bottom channel. The hole will then be bound to the electron and travel along with the electron, i.e. an exciton is formed. The third row is the single-electron transport regime ($E>V_{ij}$), in which the charge state in the drag channel is not affected by the drive-channel electron. Note that for simplicity we assume homogeneous $V_{ij} = V$, and the intra-channel and diagonal Coulomb interactions are ignored in the schematics.}
\label{CD_schematic_fig}
\end{figure*}

The experiment is carried out in an electrostatically-defined 4$\times$2 hole quantum dot array, which is fabricated in a Ge/SiGe quantum well heterostructure~\cite{Lodari2021a}. Fig.~\ref{Device_fig}(a) shows a device image, with the positions of the dots and charge sensors as indicated by the labeled circles. Fig.~\ref{Device_fig}(b) shows a schematic gate stack of the device. Screening gates, plunger gates and barrier gates were fabricated in successive lithography steps (see the Appendix for details). We refer to the path from dot 1 to dot 4 as the top channel (drive channel) and to the path from dot 5 to dot 8 as the bottom channel (drag channel). Quantum dots are formed by applying negative DC voltages on a set of plunger gates, $P$, and barrier gates, $B$, to accumulate and confine holes in the quantum well in the area between the screening gates. The charge occupation of the 4$\times$2 array is denoted $\left( \begin{smallmatrix} O_1 & O_2 & O_3 & O_4 \\ O_5 & O_6 & O_7 & O_8 \end{smallmatrix} \right)$, where $O_i$ represents the number of holes in dot $i$. The structure allows for individual control of all ten nearest-neighbor tunnel couplings. Plunger gates and barrier gates are additionally connected to high-frequency lines via bias tees to allow fast control of electrochemical potentials and tunnel couplings. 

In this experiment the plunger and barrier gates are virtualized such that changing a virtual plunger $P^\prime_i$ indepently controls the electrochemical potential, $\epsilon_i$, of dot $i$ and changing a virtual barrier $B^\prime_{ij}$ mainly modulates the tunnel coupling, $t_{ij}$, between neighboring dots $i$ and $j$ without influencing the dot potentials. In this device four charge sensors (BL, BR, TL and TR) can be formed at the four corners of the array. They serve both as detectors for the charge occupation and as reservoirs. In this experiment we use only the BL and BR sensors for charge sensing, with multiplexed RF reflectometry (TL and TR are used as reservoirs.). The plunger gates for the BL and BR sensors are also included in the gate virtualization, such that sweeping a plunger gate in the array does not shift the sensor peak position. Therefore, the sensors are mostly sensitive to changes of the charge occupation in the array. 

To study exciton formation via the Coulomb drag effect, we will aim to initialize the device in the $\left( \begin{smallmatrix} 1 & 1 & 1 & 1 \\ 0 & 0 & 0 & 0 \end{smallmatrix} \right)$ charge state, where each top-channel dot is occupied by one hole and the bottom channel is empty. Because the charge carriers in the array are holes originating from the valence band, removing a hole in the top channel amounts to adding an electron relative to the singly-filled background of holes (see Fig.~\ref{CD_schematic_fig}). We can thus load an electron to the top channel by emptying a dot (e.g. pulsing to the (0111) charge state in the top channel). The electrochemical potentials of the bottom dots in the $\left( \begin{smallmatrix} 1 & 1 & 1 & 1 \\ 0 & 0 & 0 & 0 \end{smallmatrix} \right)$ configuration are aligned with each other, such that loading a hole from the reservoir to the bottom channel costs the same energy regardless of its position. We label this energy cost $E$ (Fig.~\ref{CD_schematic_fig}). When $E$ is lower than the nearest-neighbor inter-channel Coulomb interaction $V_{ij}$, a hole will be attracted in the bottom channel by the top-channel electron, reaching e.g. the charge state $\left( \begin{smallmatrix} 0 & 1 & 1 & 1 \\ 1 & 0 & 0 & 0 \end{smallmatrix} \right)$. An electron-hole pair is thus formed bound by $V_{ij}$, which constitutes an inter-channel exciton (strictly speaking, $V_{ij}$ must here be corrected by intra-channel and diagonal Coulomb interactions; we will neglect these corrections to simplify the discussion but they are included when aligning the bottom dot potentials). Furthermore, if the system Hamiltonian favors an exciton ground state, pushing the electron (the missing hole) through the top channel will cause the hole in the bottom to move together with the electron (Fig.~\ref{Device_fig}(c) and Fig.~\ref{CD_schematic_fig}).

\section{Quantum dot ladder formation and tune-up}

\begin{figure} 
\centering    
\includegraphics[width=\columnwidth]{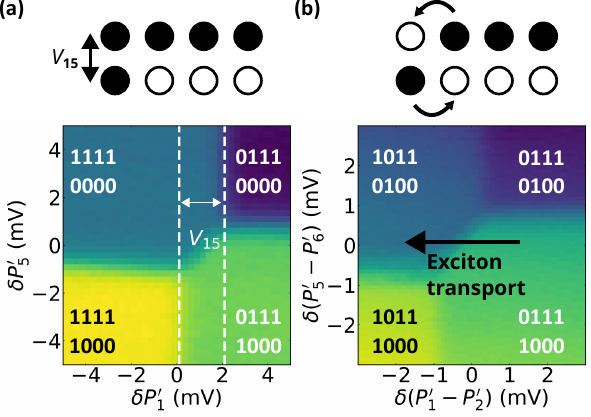}
\caption{Charge occupation control of the 4$\times$2 dot array in the single-hole regime. (a) Charge stability diagram of an inter-channel dot pair 1-5. $\delta P^{\prime}_i$ refers to the change in $P^{\prime}_i$ relative to a baseline DC voltage. The inter-channel Coulomb interaction $V_{15}$, as illustrated in the schematic above, can be determined by measuring the size of the anti-crossing (the distance between the white dashed lines) in the charge stability diagrams. The Coulomb interaction strengths are converted from voltage to energy using lever arms. See appendix for the charge-stability diagrams of all nearest-neighbor dot pairs, for all inter-channel Coulomb interaction measurements and for the extraction of lever arms. (b) Charge stability diagram as a function of 1-2 detuning $\delta(P^{\prime}_1-P^{\prime}_2)$ and 5-6 detuning $\delta(P^{\prime}_5-P^{\prime}_6)$. Along the black arrow an exciton moves in the dot array due a co-tunneling process depicted in the schematic.}
\label{Charge_tuning_fig}
\end{figure}

Figure~\ref{Charge_tuning_fig}(a) shows charge stability diagrams for the inter-channel dot pair 1-5 near the $\left( \begin{smallmatrix} 1 & 1 & 1 & 1 \\ 0 & 0 & 0 & 0 \end{smallmatrix} \right)$ charge configuration (see the Appendix for the other inter-channel pairs). The virtualized sensors result in a gradient-free signal within each charge state region. The inter-channel Coulomb interactions $V_{ij}$ between dot $i$ and $j$ are extracted from the size of the anti-crossing for an inter-dot transition. The obtained inter-channel Coulomb interaction strengths are $V_{15}$ = 220\,\si{\micro\eV}, $V_{26} = 260$\,\si{\micro\eV}, $V_{37} = 315$\,\si{\micro\eV}, and $V_{48} = 213$\,\si{\micro\eV}. The diagonal Coulomb interactions $V^\prime$ are smaller than 100\,\si{\micro\eV}.

Figure~\ref{Charge_tuning_fig}(b) shows the sensor signal as a function of the detuning of dots 1 and 2, $\delta(P^{\prime}_1-P^{\prime}_2)$, and the detuning of dots 5 and 6,  $\delta(P^{\prime}_5-P^{\prime}_6)$, near their respective inter-dot transitions. If we sweep $\delta(P^{\prime}_1-P^{\prime}_2)$ and keep $\delta(P^{\prime}_5-P^{\prime}_6)$ fixed near the 5-6 transition, as indicated by the black arrow in Fig.~\ref{Charge_tuning_fig}(b), a transition is made from $\left( \begin{smallmatrix} 0 & 1 & 1 & 1 \\ 1 & 0 & 0 & 0 \end{smallmatrix} \right)$ to $\left( \begin{smallmatrix} 1 & 0 & 1 & 1 \\ 0 & 1 & 0 & 0 \end{smallmatrix} \right)$ whereby a charge tunnels from dot 2 to dot 1 and simultaneously a charge moves from dot 5 to dot 6, thanks to the inter-channel Coulomb interactions $V_{15}$ and $V_{26}$. This co-tunneling process~\cite{Shinkai2009} results in an exciton moving in the ladder array, and is the dominant exciton transport process since it happens before sequential tunneling is energetically allowed (see the path along the black line in Fig.~\ref{Charge_tuning_fig}.

\begin{figure} 
\centering    
\includegraphics[width=\columnwidth]{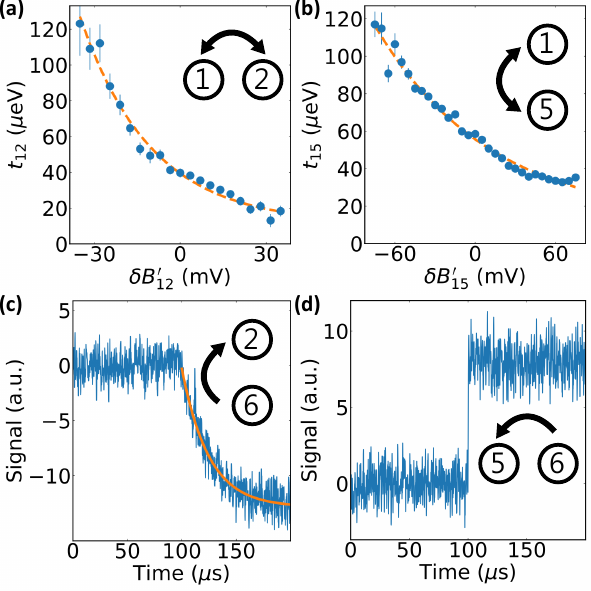}
\caption{Tunnel coupling control of the 4$\times$2 dot array in the single-hole regime. (a-b) Measurements of (a) intra-channel tunnel coupling $t_{12}$ and (b) inter-channel tunnel coupling $t_{15}$ as a function of $B^\prime_{12}$ and $B^\prime_{15}$, respectively. The orange dashed lines show exponential fits to the data. (c) Measurement of tunneling rate $\Gamma_{26}$ between dot 2 and dot 6 when $t_{26}$ is suppressed. A hole is initialized in dot 6 ($\epsilon_6 < \epsilon_2$) and at $100$\si{\micro\s} we abruptly align the electrochemical potentials ($\epsilon_6 = \epsilon_2$) using a gate voltage pulse. From this moment, the hole can tunnel from dot 6 to dot 2. The plot shows the time averaged charge sensor response. As expected, we see an exponential trend in the sensor response, since the tunnel time should obey Poisson statistics. The exponential fit yields $\Gamma_{26}=40$\,kHz, which gives a rough estimate of $t_{26}=0.03$\,\si{\micro\eV}. (d) same as (c) but for $\Gamma_{56}$ when $t_{56}=46$\,\si{\micro\eV}. $\Gamma_{56}$ is higher than the measurement bandwidth of 1\,MHz.}
\label{tc_tuning_fig}
\end{figure}

Efficient exciton transport requires strong intra-channel tunnel couplings in order to obtain large intra-channel co-tunneling couplings, and weak inter-channel tunnel couplings. Strong inter-channel tunneling exceeding the channel detuning would allow the charge carriers to hybridize between the two channels, in which case we can no longer speak of a distinct electron and hole which are bound by long-range Coulomb interaction. 

Using the gate voltages, we can control both the inter-channel and intra-channel tunnel couplings. The tunnel couplings are characterized by fitting inter-dot transition sensor signals to a model described in~\cite{VanDiepen2018}. Figures~\ref{tc_tuning_fig}(a-b) show the control of $t_{12}$ and $t_{15}$. Due to fabrication procedure, some barrier gates exhibit a weaker response than others, meaning that larger voltage swings are required for modulating the corresponding tunnel couplings (see appendix for details). Note that in the virtualized $B^\prime$ we do not compensate for tunnel coupling crosstalk~\cite{Hsiao2020b, Qiao2020} since the present experiment only requires setting the tunnel couplings once and furthermore is robust to small variations in tunnel couplings. 

We here set all intra-channel tunnel couplings to 30--40\,\si{\micro\eV}. For the inter-channel tunnel couplings we target values ideally below 1\,\si{\micro\eV}. However, it is challenging to quantify such small tunnel couplings by fitting the inter-dot sensor signal, given that the thermal energy based on the effective electron temperature is about 20\,\si{\micro\eV} in this experiment. Instead of the tunnel couplings, we measure the inter-dot tunnel rates by abruptly aligning the dot potentials using a gate voltage pulse. The relation between tunnel coupling and tunnel rate can be expressed as~\cite{Braakman2013}
\begin{equation}
\Gamma_{ij} = 2T_2t^2_{ij}
\label{tunnel_rate_formula}
\end{equation}
where $\Gamma_{ij}$ and $t_{ij}$ are the tunnel rate and tunnel couplings between dot $i$ and $j$, and $T_2$ is the charge dephasing time ($T_2 \ge$ 0.3\,ns extracted from photon-assisted-tunneling measurement~\cite{Oosterkamp1998}, see appendix for details). Figure~\ref{tc_tuning_fig}(c) shows the tunnel rate measurement between dot 2 and dot 6. The fit yields $\Gamma_{26} = 40$\,kHz. Using Eq.~(\ref{tunnel_rate_formula}) we obtain $t_{26} \le $ 0.03\,\si{\micro\eV}. For comparison, Fig.~\ref{tc_tuning_fig}(d) shows the measurement of $\Gamma_{56}$ when $t_{56} = $46\,\si{\micro\eV}. In this case the decay appears instantaneous owing to the fast tunneling between the dots. Using the inter-channel barrier voltages, all inter-channel tunnel couplings can be suppressed below 0.1\,\si{\micro\eV} (see appendix), with all inter-channel Coulomb interactions $> 150$\,\si{\micro\eV}. However, we ideally want homogeneous inter-channel Coulomb interactions of about 200-300\,\si{\micro\eV}, in order to have a large window for Coulomb drag. Since $V_{15}$ is only 166\,\si{\micro\eV} when $t_{15}=0.07$\,\si{\micro\eV}, we  bring dot 1 and dot 5 closer together to increase $V_{15}$ to 220\,\si{\micro\eV}, at the expense of a higher $t_{15}\sim25$\,\si{\micro\eV}~\footnote{We note that although $t_{15}$ is higher than other inter-channel tunnel couplings, since electron-hole pair transport is a co-tunneling process and since $t_{26}$ remains below $1 \si{\micro\eV}$, the correlated hopping of an electron-hole pair across the channels is still three orders of magnitude smaller than the hopping along the channel direction.}.

\begin{figure*} 
\centering    
\includegraphics[width=0.9\textwidth]{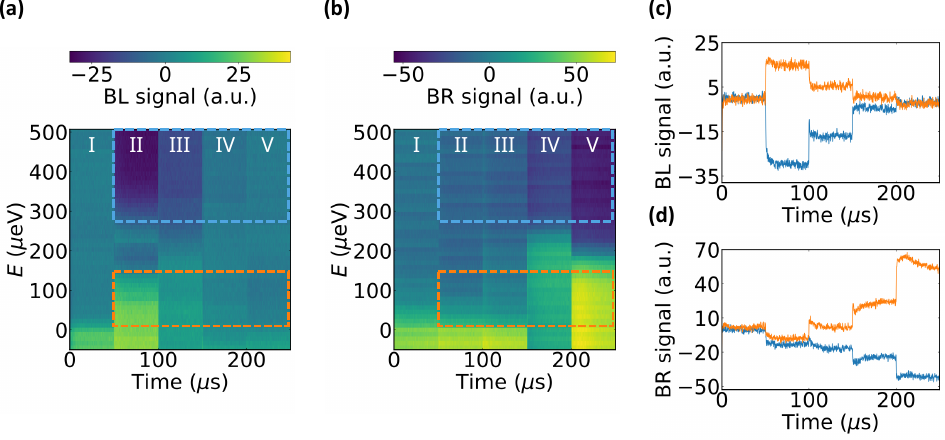}
\caption{Coulomb drag and exciton transport measurements. (a) BL and (b) BR processed (see appendix) sensor signals as a function of time and $E$. In the time domain the dot potentials in the top channel are pulsed from phase I to V as described in Fig.~\ref{CD_schematic_fig}. A positive (negative) charge causes a positive (negative) sensor signal. The regions enclosed by a blue dashed rectangle correspond to the single-electron transport regime, in which only a negative charge moves from the left to the right in the top channel. The regions enclosed by a orange dashed rectangle indicate the exciton transport regime, in which an additional positive charge is loaded and travels with the top-channel electron. Since the BL and BR sensors are more sensitive to charges in the bottom channel than in the top channel, the sensor signals change sign in the exciton transport regime compared to those in the single-electron transport regime. (c) 1D line cuts of the BL sensor signal in the single-electron transport regime ($E\simeq$500\,\si{\micro\eV}, blue trace) and in the exciton transport regime ($E\simeq$100\,\si{\micro\eV}, orange trace). (d) Same as (c) but for the BR sensor signal. The small drop in the BR signal between $50$\,$\mu$s and $100$\,$\mu$s is caused by imperfect virtualization of the BR sensor with respect to $P^{\prime}_1$. The slight bending of the BR signal is possibly caused by charging/discharging of the 2DHG near the BR sensor.}
\label{CD_data_fig}
\end{figure*}

\section{Coulomb drag and exciton formation}

The experiment scheme for measuring exciton formation and transport is illustrated in Fig.~\ref{CD_schematic_fig}. In phase I, the 4$\times$2 dot array is set to the $\left( \begin{smallmatrix} 1 & 1 & 1 & 1 \\ 0 & 0 & 0 & 0 \end{smallmatrix} \right)$ charge occupation in which the dot potentials in the top channel (drive channel) are aligned and are placed $\sim200$\,\si{\micro\eV} below the Fermi level. The potentials in the bottom channel (drag channel) are aligned as well, and positioned above the Fermi level by an energy offset, $E$. From phase II to V, the respective top-channel dot potentials are consecutively raised and then lowered by 6\,mV ($\sim670$\,\si{\micro\eV}) to load and shuttle an electron from left to right. If $E<V_{ij}$, the top-channel electron capacitively lowers the bottom-channel potential on the opposite site below the Fermi level. As a consequence a hole is loaded in the bottom channel. Due to the inter-channel Coulomb interaction, the hole is dragged along with the electron, i.e. the electron and hole move together as an exciton along the channel throughout the pulse sequence. In contrast, if $E>V_{ij}$, the top-channel electron moves alone without dragging along a hole. Therefore, a transition between exciton transport and single electron transport is expected to occur at $E \sim V_{avg} = \langle V_{ij} \rangle$. In this work, the average inter-channel Coulomb interaction $V_{avg}$ is 252\,\si{\micro\eV}. We note that for a system with inhomogeneous $V_{ij}$, the range of $E$ where Coulomb drag can occur is limited by the smallest $V_{ij}$.

In the measurements shown in Fig.~\ref{CD_data_fig}, the top-channel dot potentials are pulsed from phase I to V in the time domain while the bottom-channel potentials are fixed at $E$~\footnote{In the experiment we apply a global virtual gate voltage on the bottom channel and convert the global voltage to a global energy offset using an averaged bottom-channel lever arm 112\,\si{\micro\eV}/mV}. Figs.~\ref{CD_data_fig}(a) and (b) show the BL and BR sensor signals as a function of time and $E$. The sensor signals corresponding to the $\left( \begin{smallmatrix} 1 & 1 & 1 & 1 \\ 0 & 0 & 0 & 0 \end{smallmatrix} \right)$ charge state (phase I when $E>0$) are assigned a reference value of 0. An increasing (decreasing) signal indicates a positive (negative) charge moves closer to the corresponding sensor. In the region enclosed by the blue dashed rectangle, from phase II to V, the BL (BR) sensor signal is increasingly (less and less) negative. As $E$ is reduced, the sensor signals first pass through a transition region around $E\sim200$\,\si{\micro\eV} and then reach a region enclosed by the orange rectangle, where the BL (BR) sensor signal is less and less (increasingly) positive from phase II to V.

The data in Fig.~\ref{CD_data_fig}(a) and (b) can be understood as follows. In the blue-dashed region, the system is in the single-electron regime in which a top-channel electron is moving away from BL and towards BR. Hence, the magnitude of the negative signal decreases (increases) over time for BL (BR). In contrast, in the orange-dashed region, the system enters the exciton transport regime in which an inter-channel exciton moves to the right. Because the BL and BR sensors are more sensitive to the bottom-channel hole than to the top-channel electron, the net signal induced by the exciton is positive and the magnitude of this positive signal decreases (increases) over time for BL (BR). See Fig.~\ref{CD_data_fig}(c) and (d) for a further comparison between the signals in the single-electron-transport regime and the exciton transport regime. In Fig.~\ref{CD_data_fig}(a) and (b) the transition between the single-electron regime and the exciton transport regime occurs around $E\sim200$\,\si{\micro\eV}, which is consistent with the predicted transition point $E\sim V_{avg}=252$\,\si{\micro\eV}. The width of the transition regime depends on the level of disorder in the dot potentials ($\delta \epsilon \le 50$\,\si{\micro\eV}, which is the accuracy of the automated calibration) and variations in inter-channel $V_{ij}$ (standard deviation in $V_{ij}$ of $\sim 40$\,\si{\micro\eV}). Note that when $E<0$\,\si{\micro\eV}, the signals in phase I increase because the bottom channel starts loading holes from the reservoirs, even if no electron is loaded in the top channel.

Finally, since the transport of an inter-channel exciton involves a co-tunneling process, it is possible in principle that either the electron or the hole or the entire exciton are not successfully transferred from one site to another. In the data of Fig.~\ref{CD_data_fig}, no such failed charge transfers are observed. This is expected since the 50\,\si{\micro\s} duration of the pulse segments by far exceeds both the single-particle tunneling rates and the co-tunneling rates (in the Appendix, we estimate the probability of successful adiabatic charge transfer to be about $99.2\%$). 

\section{Conclusion and outlook}

In summary, we have fabricated a germanium 4$\times$2 quantum dot ladder and use it as a quantum simulator for exciton formation. To engineer the simulator Hamiltonian, we tune the full array into the single-hole regime and independently control all the on-site potentials and interdot tunnel couplings. We find strong inter-channel Coulomb interaction while the tunneling between channels is suppressed, which is essential for simulating excitonic physics. To probe exciton formation by means of Coulomb drag, we drive an electron through the top channel and measure the charge sensor signals as a function of the bottom channel potential. The measured signals are in good agreement with the picture of a transition from single-electron transport to  exciton transport resulting from the inter-channel Coulomb interaction. An interesting next step possible with the present sample is to create and study an engineered excitonic insulator~\cite{Jerome1967}. 

In the future, we envision that with sufficiently homogeneous interaction energies and co-tunnel couplings in longer ladders, excitons can delocalize over the array, show coherent dynamics in the time domain, and exciton quasi-condensation~\footnote{Strictly speaking, exciton condensation does not occur in 1D or 2D at finite temperature. However, for real experimental systems we can have quasi-condensation when the correlation length exceeds the system size~\cite{Petrov2000}}.
It is useful to point out an enhanced symmetry in bilinear quantum dot arrays as described by Eq.~\ref{H_ex_eq}, which should play an important role in the nature of the ground state in the thermodynamic limit. As there is no tunnelling between the channels, one can define separate SU(2) symmetries for each channel~\footnote{Holes in strained germanium have spin-3/2, but the large heavy-hole light-hole splitting leads to an effective two-level system.}. The full Hamiltonian is symmetric with respect to both of them, and the full symmetry of the system is SO(4) $\simeq$ SU(2) $\otimes$ SU(2)\cite{YANG1990}. Excitonic condensation in this system would require spontaneous symmetry breaking of the SO(4) symmetry. For non-Abelian symmetries such as SO(4), the Hohenberg-Mermin-Wagner theorem shows that only exponentially decaying correlations are allowed even at zero temperature, due to the abundance of possible fluctuations of the order parameter. Interestingly, two excitons can together form a SO(4) singlet. Such singlets can exhibit quasi-long range order at zero temperature in one dimensional systems, analogously to spinless bosons. This suggests our system can exhibit unusual types of ground states in the thermodynamic limit, such as quasi-condensates of composite particles or states with broken translational symmetry. Analogues phenomena have been discussed in the context of spinor condensates of cold atoms in one-dimensional systems\cite{Rizzi2005,Shlyapnikov2011}.

One can also break the SO(4) symmetry by introducing extra terms to the Hamiltonian. When breaking SO(4) symmetry with a magnetic field, $S_z=1$ excitons are favored and can form a (quasi-)condensate, which is not usually seen in optical spectroscopy since these excitons are dark. Finally, the spin-orbit coupling present in germanium quantum wells, while not breaking time reversal symmetry~\cite{Winkler2003}, can also hybridize singlet and triplet states, lifting their degeneracy~\cite{Gorkov2001,Golovach2008}, which may lead to condensation at zero magnetic field.

\begin{acknowledgments}
We acknowledge useful discussions with members of the Vandersypen group, and with D. Sels, S. Gopalakrishnan, A. Bohrdt, F. Grusdt, I. Morera, H. Lange. We thank software development by S. L. de Snoo. We also acknowledge technical support by O. Benningshof, J. D. Mensingh, R. Schouten, E. Van der Wiel and N. P. Alberts. L.M.K.V. acknowledges support from an Advanced Grant of the European Research Council (ERC) under the European Union’s Horizon 2020 research and innovation programme (grant agreement No 882848) and by a Vici grant of the Dutch Research Council (NWO). E.D. acknowledges support from the ARO grant number W911NF-20-1-0163, and the SNSF project 200021-212899. 
\end{acknowledgments}
    
\section*{Data availability}
The data reported in this paper are archived on a Zenodo data repository at https://doi.org/10.5281/zenodo.8105397

\begin{appendix}

\subsection{Device fabrication and experiment setup}

The device was fabricated on a Ge/SiGe heterostructure featuring a strained Ge quantum well positioned 55\,nm below the semiconductor-dielectric interface, as described in~\cite{Lodari2021a}. The fabrication started by defining ohmic contacts, which were made by electron beam lithography, etching of the native oxide with buffered HF, and electron beam deposition of 30\,nm of Al. An insulating layer of 7\,nm Al\textsubscript{2}O\textsubscript{3} was grown with atomic layer deposition, also annealing the device and diffusing the aluminum into the heterostructure during the process. Subsequently, the screening gates (3/17\,nm Ti/Pd), plunger gates (3/27\,nm Ti/Pd), and barrier gates (3/37\,nm Ti/Pd) were made in three metalization layers, which are all separated by 5\,nm thick layers of Al\textsubscript{2}O\textsubscript{3}. Note that for easing the lift-off of the compact barrier gates, we made the barrier gates in two steps, in which the barrier gates were distributed in two lithography/evaporation/lift-off processes without a Al\textsubscript{2}O\textsubscript{3} layer in between.

The measurement was performed in a Oxford Instruments Triton dilution refrigerator with a nominal base temperature of 6\,mK. The device was mounted on a custom-made sample PCB. DC voltages from homebuilt SPI DAC modules and pulses from a Keysight M3202A AWG are combined using on-PCB bias-tees. RF reflectometry for charge sensing was done using SPI IQ-demodulation modules and on-PCB LC tank circuits. The demodulated signals were recorded by a Keysight M3102A digitizer.

\subsection{Single-hole regime of the 4$\times$2 array}

\begin{figure*}
\centering    
\includegraphics[width=\linewidth]{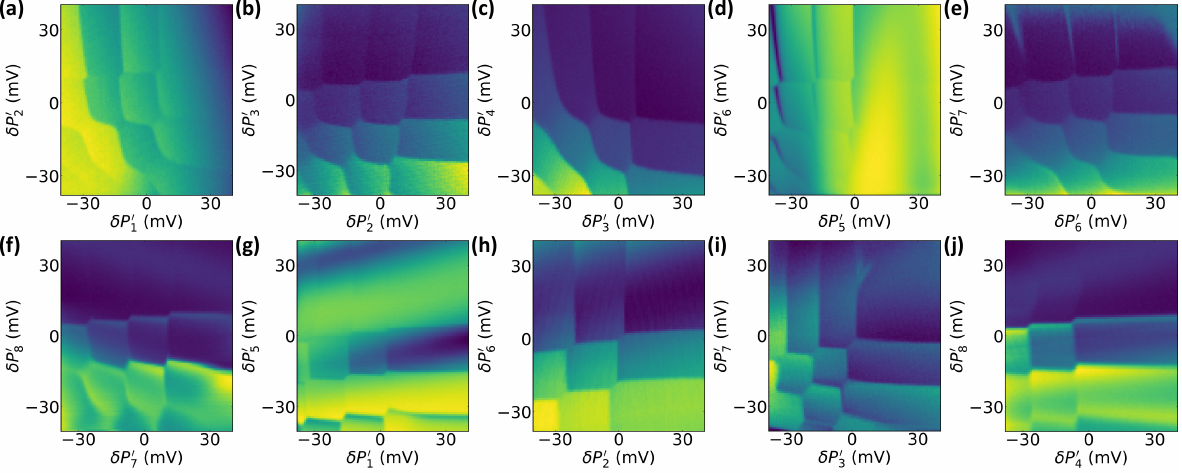}
\caption{Charge stability diagrams of all neighbouring dot pairs, showing the left or right bottom sensor signal as a function of two virtual plunger gate voltages. (a)-(f) Dot pairs along the top or bottom channel. (g)-(j) Dot pairs across the channel. The top right corner of each figure corresponds to the zero-charge state.}
\label{CSD_all_fig}
\end{figure*}

\begin{figure}
\centering    
\includegraphics[width=0.56\columnwidth]{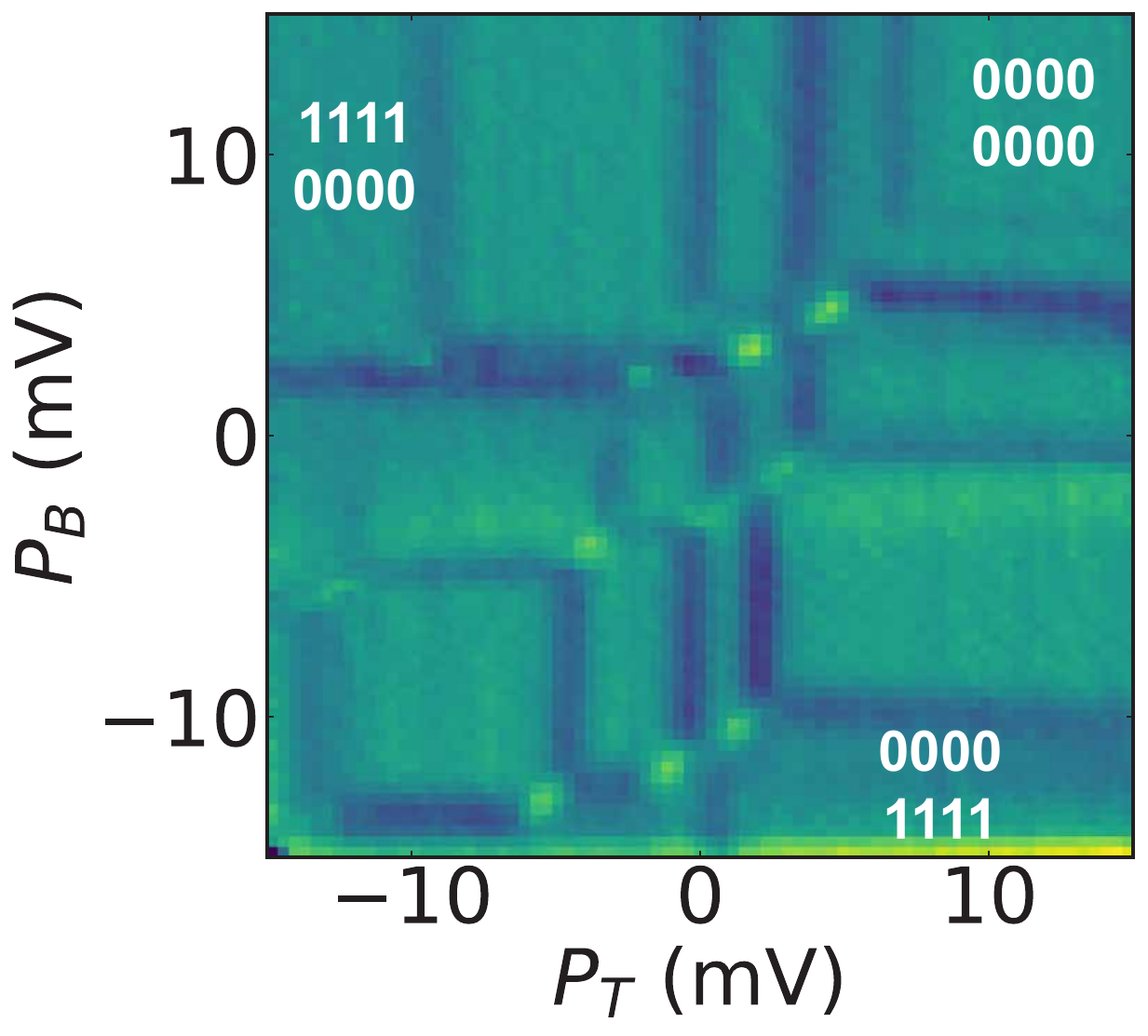}
\caption{Charge stability diagram depicting the global charge state tunability of the 4x2 array. The scanned gates are the top and bottom channel virtual voltage $P_T = P^{\prime}_1 + P^{\prime}_2 + P^{\prime}_3 + P^{\prime}_4$ and $P_B = P^{\prime}_5 + P^{\prime}_6 + P^{\prime}_7 + P^{\prime}_8$. Every vertical (horizontal) charge transition corresponds to adding a single charge to the top (bottom) channel. We plot the gradient of the sensor signal for better resolution of the transition lines. The top left region of this plot corresponds to a filled top channel and an empty bottom channel, which is the starting charge configuration for the Coulomb drag experiment.}
\label{CSD_gradient_fig}
\end{figure}

The charge state tunability of the 4$\times$2 ladder is displayed in Fig.~\ref{CSD_all_fig}, where we show charge stability diagrams for all dot pairs down to the single-hole regime. The area on the top right corner of the plots corresponds to the zero-charge state. The effect of gate voltage crosstalk is compensated using virtual gates $P^{\prime}$. All ten plots are obtained using charge sensing using the bottom right and bottom left sensors.

Additionally, in Fig.~\ref{CSD_gradient_fig}, we show global charge state control of full the 4$\times$2 array by sweeping two virtual gates, corresponding to the top and bottom channel energies ($P_T$ and $P_B$, respectively). Every vertical or horizontal addition line reflects a single charge being added to either the top or the bottom channel. Lines are spaced apart by the long-range Coulomb interaction. The starting charge occupation for the Coulomb drag experiment corresponds to the top left of this plot, with 4 charges in the top channel and none in the bottom.

\subsection{Lever arm measurement}

\begin{figure}
\centering    
\includegraphics[width=\columnwidth]{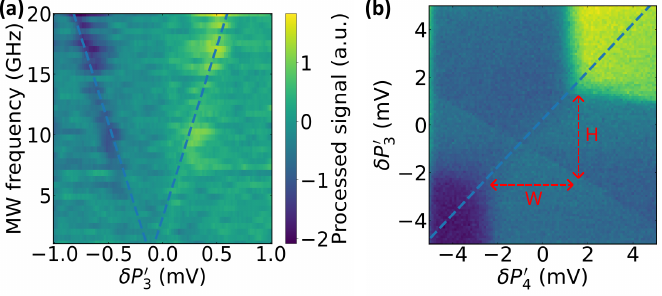}
\caption{Example of lever arm measurements. (a) PAT measurement showing the processed sensor signal as a function of frequency and detuning at the inter-dot transition between dot 3 and dot 7. The blue dashed line is a fit of the form $hf = \sqrt{\delta\epsilon_{3}^2+4t_{37}^2}$, where $\delta\epsilon_{3}=L_{3}\delta P_{3}^\prime$. The linewidth is about 0.1\,mV (11.7\,\si{\micro\eV}), from which we estimate a charge coherence time $T_2 \ge 0.3$\,ns. (b) The charge stability diagram at the inter-dot transition between dot 3 and dot 4. The blue dashed line shows the inter-dot transition line with a slope $S = H/W = L_4/L_3$ where $H$ is the height and $W$ is the width of the transition line.}
\label{lever_arm_fig}
\end{figure}

The conversion between a virtual gate voltage $P^{\prime}_{i}$ and electrochemical potential $\epsilon_{i}$ is described by $\delta\epsilon_{i}=L_i\delta P^{\prime}_{i}$, where $L_{i}$ is the lever arm for dot $i$. The lever arms can be characterized using photon-assisted tunneling (PAT)~\cite{Oosterkamp1998}. In Fig.~\ref{lever_arm_fig}(a), the signal is fitted to $hf = \sqrt{\delta\epsilon_{3}^2+4t_{37}^2}$. From the fit a lever arm $L_{3} = 117$\,\si{\micro\eV}/mV is extracted. In addition, the ratio between two lever arms can be determined from the slope, $S$, of an inter-dot charge transition line based on the fact that $V_{ij}=V_{ji}$. For instance, in Fig.~\ref{lever_arm_fig}(b), $V_{34}=L_3 H = V_{43} = L_4 W$. Therefore, $S = H/W = L_4/L_3$. So, $L_4$ can be estimated from $L_3$ and $S$. We obtain $L_4=117$\,\si{\micro\eV}/mV with $L_3=117$\,\si{\micro\eV}/mV and $S=$1.0. Similarly, based on PAT measurements and inter-dot slopes, all lever arms are estimated. The results are summarized in table~\ref{summarized_leverarms}. All lever arms have similar value $\sim 110$\,\si{\micro\eV}/mV with a standard deviation of $4$\,\si{\micro\eV}/mV.

\begin{table}
\centering
\begin{tabular}{ |c |c | c |}
 \hline
  $L_i$ & Value (\si{\micro\eV}/mV) & Method\\
  \hline
  $L_1$ & 111 & Inter-dot slope \\
  $L_2$ & 104 & PAT \\
  $L_3$ & 117 & PAT \\
  $L_4$ & 117 & Inter-dot slope \\
  $L_5$ & 115 & Inter-dot slope \\
  $L_6$ & 113 & PAT \\
  $L_7$ & 112 & PAT \\
  $L_8$ & 111 & Inter-dot slope \\
 \hline
\end{tabular}
\caption{The values and measurement methods for each lever arm $L_{i}$.}
\label{summarized_leverarms}
\end{table}

\subsection{Inter-channel Coulomb interaction measurement}

\begin{figure}
\centering    
\includegraphics[width=\columnwidth]{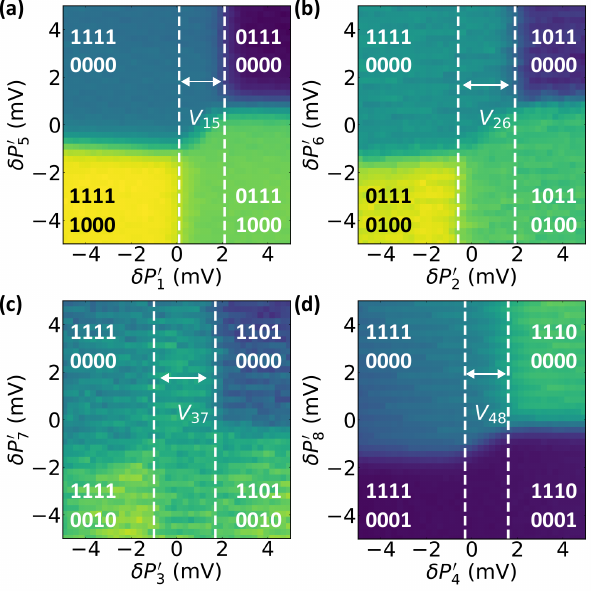}
\caption{Measurements of inter-channel Coulomb interactions (a) $V_{15}$ (replotted from Fig.~\ref{Charge_tuning_fig}a), (b) $V_{26}$, (c) $V_{37}$ and (d) $V_{48}$.}
\label{Vij_measurement_fig}
\end{figure}

Figure~\ref{Vij_measurement_fig}(a)-(d) shows the measurements of inter-channel Coulomb interactions, which are responsible for the excitonic Coulomb drag effect. As in Figure~\ref{Charge_tuning_fig}(a), the Coulomb interactions are characterized by finding the sizes of the anti-crossings and converting them into energies through lever arms. From Fig.~\ref{Vij_measurement_fig}(a)-(d) we obtain $V_{15}=220$\,\si{\micro\eV}, $V_{26}=260$\,\si{\micro\eV}, $V_{37}=315$\,\si{\micro\eV}, and $V_{48}=213$\,\si{\micro\eV}.

\subsection{Tunnel coupling control}

\begin{figure*}
\centering    
\includegraphics[width=\linewidth]{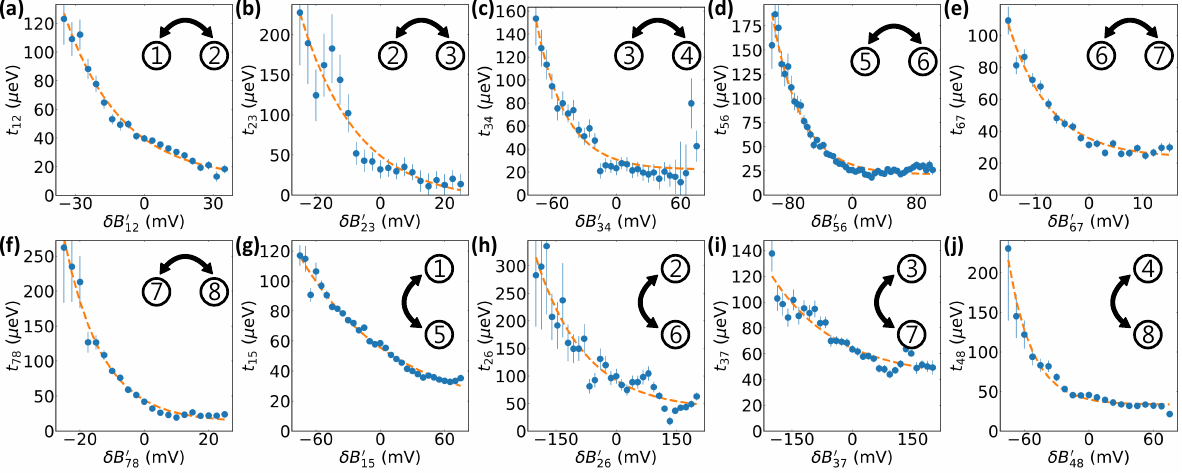}
\caption{Control of all nearest-neighbour tunnel couplings. (a)-(f) Measured intra-channel tunnel coupling $t_{ij}$ as a function of virtual barrier gates $B^{\prime}_{ij}$. (g)-(j) Measured inter-channel tunnel coupling $t_{ij}$ as a function of virtual barrier gates $B^{\prime}_{ij}$. The orange dashed lines are exponential fits to the data.}
\label{all_tc}
\end{figure*}

Figure~\ref{all_tc} shows control of all nearest-neighbour tunnel couplings $t_{ij}$ using the corresponding virtual barrier gates $B^{\prime}_{ij}$. The tunnel coupling dependency is fitted by an exponential function $A \exp(-\gamma_{ij}B^{\prime}_{ij}) + C$, from which the barrier lever arm $\gamma_{ij}$ is extracted. The $\gamma_{ij}$ are summarized in table~\ref{summarized_barrier_leverarms}. Roughly, the barrier lever arms can be separated into two groups, corresponding to the two steps in which the barriers were fabricated. Notably, the barrier gates patterned in the first fabrication step display a stronger lever arm than those patterned in the second step, despite the absence of an ALD layer between the two barrier metalization layers. The reasons for this discrepancy requires further investigation, but might be caused by the device design or residual resist under the second barrier gate layer. Nonetheless, all barriers display a reasonable level of tunnel coupling control, which allows us to tune the tunnel couplings to the values required to perform the excitonic Coulomb drag experiment.

\begin{table}
\centering
\begin{tabular}{ |c |c | c |}
 \hline
  $\gamma_{ij}$ & Value (1/mV) & Barrier fabrication step\\
  \hline
  $\gamma_{12}$ & 0.040 & 1st \\
  $\gamma_{23}$ & 0.057 & 1st \\
  $\gamma_{34}$ & 0.036 & 2nd \\
  $\gamma_{56}$ & 0.028 & 2nd \\
  $\gamma_{67}$ & 0.128 & 1st \\
  $\gamma_{78}$ & 0.085 & 1st \\
  $\gamma_{15}$ & 0.012 & 1st \\
  $\gamma_{26}$ & 0.008 & 2nd \\
  $\gamma_{37}$ & 0.006 & 2nd \\
  $\gamma_{48}$ & 0.044 & 1st \\
 \hline
\end{tabular}
\caption{The values and corresponding barrier fabrication steps for each tunnel barrier lever arm $\gamma_{ij}$.}
\label{summarized_barrier_leverarms}
\end{table}

\subsection{Tunnel rate measurement}

\begin{figure*}
\centering    
\includegraphics[width=0.9\linewidth]{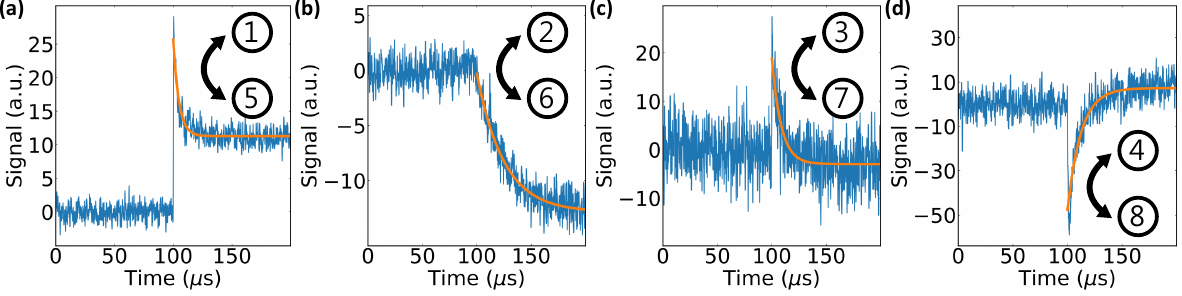}
\caption{Tunnel rate measurements for (a) $\Gamma_{15}$, (b)  $\Gamma_{26}$, (c)  $\Gamma_{37}$, and (d)  $\Gamma_{48}$. The orange curves are exponential fits to the data.}
\label{all_tunnel_rate}
\end{figure*}

Tunnel coupling extraction via fitting of the inter-dot transition signals allows us to reliably obtain tunnel coupling values of the order of tens of \si{\micro\eV}, larger than or comparable to the electron temperature. As $t_{ij}$ becomes much smaller than the electron temperature, this fit becomes unreliable. When the hopping between channels is suppressed, we estimate the inter-channel $t_{ij}$ from the inter-channel tunnel rates $\Gamma_{ij}$ as described in the main text. Figure~\ref{all_tunnel_rate} (a)-(d) show the tunnel rate measurements, from which we obtain $\Gamma_{15}=208$\,kHz, $\Gamma_{26}=40$\,kHz, $\Gamma_{37}=118$\,kHz, and $\Gamma_{48}=81$\,kHz. Since we estimate $T_2 \ge 0.3$\,ns (lower limit) from the linewidth of the PAT in Fig.~\ref{lever_arm_fig}(a), by using Eq.~\ref{tunnel_rate_formula} we can then estimate $t_{15}\le0.07$\,\si{\micro\eV}, $t_{26}\le0.03$\,\si{\micro\eV}, $t_{37}\le0.06$\,\si{\micro\eV}, and $t_{48}\le0.05$\,\si{\micro\eV} in the target regime where the inter-channel hopping is suppressed.

\subsection{Automated calibration routine}

\begin{figure*}
\centering    
\includegraphics[width=\linewidth]{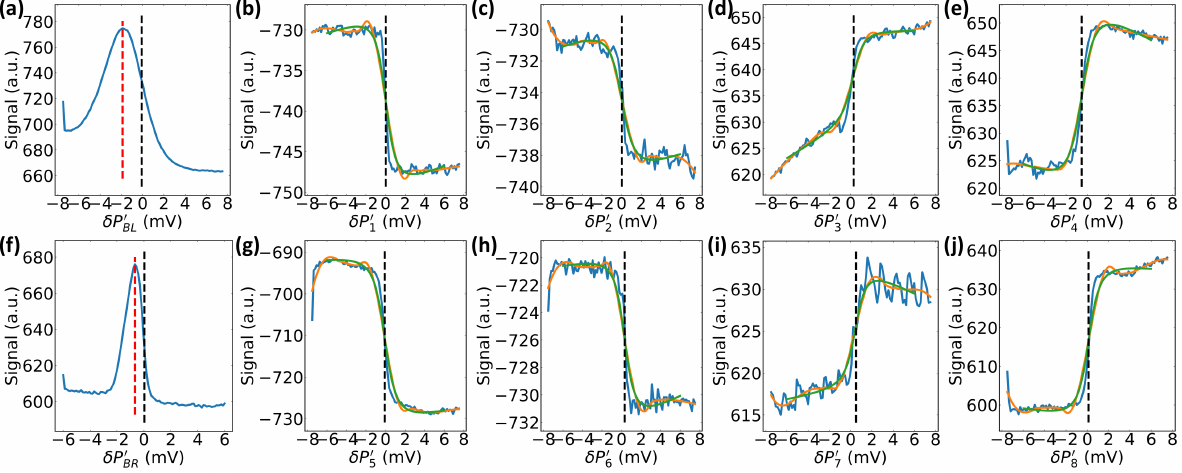}
\caption{Fast automated calibration routine. (a) and (f) show the sensor signals as a function of sensor plunger gates $P^{\prime}_{BL}$ (BL sensor) and $P^{\prime}_{BL}$ (BR sensor). The red dashed lines indicate the sensor peak positions and the black dashed lines indicate the optimal sensing positions (where the slope is steepest). $P^{\prime}_{BL}$ and $P^{\prime}_{BL}$ are subsequently moved to the optimal sensing positions. (b)-(e) and (g)-(j) are the sensor signals for dot 1 to 8, near the addition lines for the first hole of each dot. The black dashed lines show the voltages at which $\epsilon_i$ aligns with the Fermi level. Note that (b)-(e) and (g)-(j) are not centered at the same DC voltages. Instead, each addition line is taken with an added offset $\Delta P^{\prime}_{i, target}$ from the original DC voltage $V_{base}$. $V_{base}$ is then adjusted according to the deviation of the black dashed lines from the centers of the scans. Each scan takes approximately 1 second and the whole calibration routine takes about 10 seconds.}
\label{auto_calibration_fig}
\end{figure*}

Slow changes in the electrostatic environment of the device lead to inevitable drift of dot electrochemical potentials. To compensate for this low frequency drift, we implement a fast automated calibration routine to keep the electrochemical potentials fixed relative to the Fermi level. Our target is to set the level of dot $i$ with an offset $P^{\prime}_{i, target}$ from the Fermi level. In this experiment $P^{\prime}_{1, target}$ to $P^{\prime}_{8, target}$ are initially [2, 2, 2, 2, -4, -4, -4, -4]\,mV, which places the device in the $\left( \begin{smallmatrix} 1 & 1 & 1 & 1 \\ 0 & 0 & 0 & 0 \end{smallmatrix} \right)$ charge state. For the first instance of the calibration, we manually tune the device to a baseline DC voltage $V_{base}$ close to the target condition (within a tolerance of a few mV). The calibration routine starts with optimizing the sensor signals, which is done by scanning sensor plunger gates and locating the optimal sensing positions, as shown in Fig.~\ref{auto_calibration_fig}(a) and (f). The voltage drift of dot $i$ is measured by scanning $P^{\prime}_{i}$ centered at $V_{base} + P^{\prime}_{i, target}$ and fitting the signal to a charge addition line to locate the Fermi level, as shown in Fig.~\ref{auto_calibration_fig} (b)-(e) and (g)-(j). $V_{base}$ is subsequently shifted by the deviation of the addition lines from the centers of the scans to compensate for the voltage drift. The entire automated calibration routine takes about 10 seconds and offers a valuable tool for the efficient adjustment of dot potentials in multi-dot devices.

\subsection{Exciton tunnel coupling}

The tunneling of excitons entails a co-tunneling process of two charges in the ladder array. Here we take the tunneling between $\left( \begin{smallmatrix} 1 & 0 & 1 & 1 \\ 0 & 1 & 0 & 0 \end{smallmatrix} \right)$ and $\left( \begin{smallmatrix} 1 & 1 & 0 & 1 \\ 0 & 0 & 1 & 0 \end{smallmatrix} \right)$ as an example. The relevant charge states are $\Ket{0} =\left( \begin{smallmatrix} 1 & 0 & 1 & 1 \\ 0 & 1 & 0 & 0 \end{smallmatrix} \right)$, $\Ket{1} =\left( \begin{smallmatrix} 1 & 1 & 0 & 1 \\ 0 & 0 & 1 & 0 \end{smallmatrix} \right)$, $\Ket{2} =\left( \begin{smallmatrix} 1 & 1 & 0 & 1 \\ 0 & 1 & 0 & 0 \end{smallmatrix} \right)$, $\Ket{3} =\left( \begin{smallmatrix} 1 & 0 & 1 & 1 \\ 0 & 0 & 1 & 0 \end{smallmatrix} \right)$, $\Ket{4} =\left( \begin{smallmatrix} 1 & 1 & 1 & 1 \\ 0 & 0 & 0 & 0 \end{smallmatrix} \right)$, and $\Ket{5} =\left( \begin{smallmatrix} 1 & 0 & 0 & 1 \\ 0 & 1 & 1 & 0 \end{smallmatrix} \right)$. The Hamiltonian in this basis is

\begin{equation}
H=
\begin{pmatrix}
    E_{0} & 0 & -t_{23} & -t_{67} & -t_{26} & -t_{37}\\
    0 & E_{1} & -t_{67} & -t_{23} & t_{37} & t_{26}\\
    -t_{23} & -t_{67} & E_{2} & 0 & 0 & 0\\
    -t_{67} & -t_{23} & 0 & E_{3} & 0 & 0\\
    -t_{26} & t_{37} & 0 & 0 & E_{4} & 0\\
    -t_{37} & t_{26} & 0 & 0 & 0 & E_{5}\\
\end{pmatrix}    
\label{H_origin}
\end{equation}

where

\begin{equation}
\begin{aligned}
&E_{0} = -\epsilon_{3}-\epsilon_{6}+V+2V^{\prime}-\epsilon_{1}-\epsilon_{4}\\
&E_{1} = -\epsilon_{2}-\epsilon_{7}+V+2V^{\prime}-\epsilon_{1}-\epsilon_{4}\\
&E_{2} = -\epsilon_{2}-\epsilon_{6}+2V+V^{\prime}-\epsilon_{1}-\epsilon_{4}\\
&E_{3} = -\epsilon_{3}-\epsilon_{7}+2V+V^{\prime}-\epsilon_{1}-\epsilon_{4}\\
&E_{4} = -\epsilon_{2}-\epsilon_{3}+3V-\epsilon_{1}-\epsilon_{4}\\
&E_{5} = -\epsilon_{6}-\epsilon_{7}+V+2V^{\prime}-\epsilon_{1}-\epsilon_{4}\\
\end{aligned}
\label{energy_eq_1}
\end{equation}

$V$ the nearest-neighbor Coulomb interaction and $V^{\prime}$ the diagonal Coulomb interaction (for simplicity we assume homogeneous $V$ and $V^{\prime}$ in the ladder array). Near a symmetric exciton tunneling condition in which $\epsilon_{2}\approx\epsilon_{3}=\epsilon+\Delta$, $\epsilon_{6}\approx\epsilon_{7}=\epsilon$, and $(V-V^{\prime}), \Delta \gg t_{23}, t_{67}, t_{26}, t_{37}$, Eq.~\ref{energy_eq_1} becomes

\begin{equation}
\begin{aligned}
&E_{0} = -2\epsilon-\Delta+V+2V^{\prime}+\delta E_{0}\\
&E_{1} = -2\epsilon-\Delta+V+2V^{\prime}+\delta E_{1}\\
&E_{2} = -2\epsilon-\Delta+2V+V^{\prime}+\delta E_{2}\\
&E_{3} = -2\epsilon-\Delta+2V+V^{\prime}+\delta E_{3}\\
&E_{4} = -2\epsilon-2\Delta+3V+\delta E_{4}\\
&E_{5} = -2\epsilon+V+2V^{\prime}+\delta E_{5}\\
\end{aligned}
\label{energy_eq_2}
\end{equation}

where $\delta E_{i}$ is a small perturbation of $E_{i}$ near the symmetric exciton tunneling condition. We then express Eq.~\ref{H_origin} in the eigenbasis of the first-order perturbation $H^{\prime} \simeq U^{\dagger}HU$ in which

\begin{widetext}
\begin{equation}
U=
\begin{pmatrix}
    1 & 0 & -\frac{t_{23}}{V-V^{\prime}} & -\frac{t_{67}}{V-V^{\prime}} & -\frac{t_{26}}{2V-2V^{\prime}-\Delta} & -\frac{t_{37}}{\Delta}\\
    0 & 1 & -\frac{t_{67}}{V-V^{\prime}} & -\frac{t_{23}}{V-V^{\prime}} & \frac{t_{37}}{2V-2V^{\prime}-\Delta} & \frac{t_{26}}{\Delta}\\
    \frac{t_{23}}{V-V^{\prime}} & \frac{t_{67}}{V-V^{\prime}} & 1 & 0 & 0 & 0\\
    \frac{t_{67}}{V-V^{\prime}} & \frac{t_{23}}{V-V^{\prime}} & 0 & 1 & 0 & 0\\
    \frac{t_{26}}{2V-2V^{\prime}-\Delta} & -\frac{t_{37}}{2V-2V^{\prime}-\Delta} & 0 & 0 & 1 & 0\\
    \frac{t_{37}}{\Delta} & -\frac{t_{26}}{\Delta} & 0 & 0 & 0 & 1\\
\end{pmatrix} 
\label{H_purturbed}
\end{equation}
\end{widetext}

Neglecting terms of more than second order in $\frac{t_{ij}}{V-V^{\prime}}$, $\frac{t_{ij}}{2V-2V^{\prime}-\Delta}$ or $\frac{t_{ij}}{\Delta}$ the effective Hamiltonian $H^{\prime}$ for the perturbed states $\Ket{0'}$ and $\Ket{1'}$ becomes

\begin{equation}
H^{\prime}=
\begin{pmatrix}
    E_{0'} & -t_{co}\\
    -t_{co} & E_{1'}\\
\end{pmatrix}   
\label{H_eff}
\end{equation}

where $E_{0'} = E_{0}-\frac{t_{23}^2}{V-V^{\prime}}-\frac{t_{67}^2}{V-V^{\prime}}-\frac{t_{26}^2}{2V-2V^{\prime}-\Delta}-\frac{t_{37}^2}{\Delta}$, $E_{1'} = E_{1}-\frac{t_{23}^2}{V-V^{\prime}}-\frac{t_{67}^2}{V-V^{\prime}}-\frac{t_{37}^2}{2V-2V^{\prime}-\Delta}-\frac{t_{26}^2}{\Delta}$, and $t_{co} = 2\frac{t_{23}t_{67}}{V-V^{\prime}}-\frac{t_{26}t_{37}}{2V-2V^{\prime}-\Delta}-\frac{t_{26}t_{37}}{\Delta}$~\cite{Braakman2013}. From Eq.~\ref{H_eff} we see that the tunneling of exciton states is determined by $t_{co}$, which has a term proportional to the product of intra-channel tunnel couplings and a term proportional to the product of inter-channel tunnel couplings. In the present experiment, the former is much larger than the latter by at least three orders of magnitude. Therefore, $t_{co}$ is predominantly caused by the co-tunneling of charges in the intra-channel direction.

\subsection{Adiabatic exciton transfer probability}

We estimate the probability that an exciton adiabatically transitions between neighbouring sites in the quantum dot ladder. When this transition does not occur adiabatically, the exciton initially stays where it was. Afterwards, either the electron or the hole may tunnel, leaving the other particles behind, and eventually the entire exciton may still transition, but at least for a brief moment in time the intended exciton transport does not take place. For instance, the transition from $\left( \begin{smallmatrix} 1 & 0 & 1 & 1 \\ 0 & 1 & 0 & 0 \end{smallmatrix} \right)$  to $\left( \begin{smallmatrix} 1 & 1 & 0 & 1 \\ 0 & 0 & 1 & 0 \end{smallmatrix} \right)$ might instead end with $\left( \begin{smallmatrix} 1 & 0 & 1 & 1 \\ 0 & 1 & 0 & 0 \end{smallmatrix} \right)$ (the pair is not transferred) or $\left( \begin{smallmatrix} 1 & 1 & 0 & 1 \\ 0 & 1 & 0 & 0 \end{smallmatrix} \right)$ (a hole lags behind). Using the Landau-Zener formula~\cite{L.D.Landau1932,Zener1932}, we obtain
\begin{equation}
\begin{aligned}
&P_{dia}&=&\quad \exp\left(-2\pi\frac{t_{co}^2}{\hbar V_E}\right)\\
&t_{co}&=&\quad \frac{2t^2}{V-V^\prime}\\
&V_E&=&\quad \frac{\Delta E}{\Delta T_r}\\
\end{aligned}
\label{LZ_formula}
\end{equation}
where $P_{dia}$ is the diabatic transition probability for the transition from $\left( \begin{smallmatrix} 1 & 0 & 1 & 1 \\ 0 & 1 & 0 & 0 \end{smallmatrix} \right)$  to $\left( \begin{smallmatrix} 1 & 1 & 0 & 1 \\ 0 & 0 & 1 & 0 \end{smallmatrix} \right)$, $t_{co}$ is the intra-channel co-tunneling of the electron-hole pair, $V_E$ is the energy level velocity, $t$ is the intra-channel tunnel coupling, $V$ is the inter-channel Coulomb interaction, $V^\prime$ is the diagonal Coulomb interaction, $\Delta E$ is the energy difference between the $\left( \begin{smallmatrix} 1 & 0 & 1 & 1 \\ 0 & 1 & 0 & 0 \end{smallmatrix} \right)$  and $\left( \begin{smallmatrix} 1 & 1 & 0 & 1 \\ 0 & 0 & 1 & 0 \end{smallmatrix} \right)$ charge states, and $\Delta T_r$ is the rise time of the pulse. Note that we do not include the inter-channel co-tunneling processes in the analysis because they are at least three orders of magnitude smaller than that of the intra-channel co-tunneling process, as discussed before. Entering the experimental parameters, we obtain $P_{dia}\simeq0.8\%$. Therefore, during Coulomb drag the inter-channel exciton is transported adiabatically with an estimated fidelity of $99.2\%$.

\subsection{Coulomb drag data processing}

In Fig.~\ref{CD_data_fig} the raw data of the BR sensor signal is inverted such that an increasing (decreasing) signal corresponds to a positive (negative) charge. In addition, residual crosstalk from the bottom virtual gates to the sensor signals leads to a small gradient along $E$ (y axis) in phase I of Fig.~\ref{CD_data_fig}(a) and (b). We remove this residual crosstalk by fitting the signals in phase I to a linear background signal and subtracting this background from the data of the entire panel. The scripts for data processing can be found in the data repository.

\subsection{Numerical simulation of exciton transport}

We perform numerical simulations to compare with the measured exciton transport data in Fig. \ref{CD_data_fig}(a) and (b). To this end, we compute the ground state charge configuration of a classical Fermi-Hubbard Hamiltonian:

\begin{equation}
\begin{aligned}
H = &\sum_{i}\epsilon_in_i  + U\sum_{i}\frac{n_{i}(n_{i}-1)}{2} + U^{\prime}\sum_{\langle i,j \rangle}n_{i}n_{j}\\ &+ \sum_{i\in \alpha,j\in \beta}V_{ij}n_{i}n_{j} + V^{\prime}\sum_{i\in \alpha,j\in \beta}n_{i}n_{j}\\
\end{aligned}
\label{H_ex_clas}
\end{equation}

Compared to Eq.~\ref{H_ex_eq}, we have set $t = 0$ to facilitate the computation. We further include the electrochemical potentials $\{\epsilon_i\}$ and account for the experimentally observed differences in inter-channel Coulomb repulsion $V_{ij}$.

Due to the absence of tunnel coupling terms, this simple Hamiltonian is already diagonal. Finding its ground state charge configuration becomes therefore a straight-forward energy minimization problem. Since $U \gg V_{ij}, U^{\prime}$, double occupations are always high in energy and it suffices to input homogeneous charging energies $U \sim \SI{2}{meV}$ as extracted from charge stability diagrams in Fig. \ref{CSD_all_fig}. To capture the observed variations of the exciton transport windows, it is necessary to input the measured inter-channel Coulomb interaction parameters $V_{ij}$ as specified in the main text (see section IV, neglecting diagonal interactions). Furthermore, for the intra-channel Coulomb interaction, we assume homogeneous interaction terms $U^{\prime} \sim \SI{400}{\micro eV}$. 

\begin{figure*}
\centering    
\includegraphics[width=0.8\linewidth]{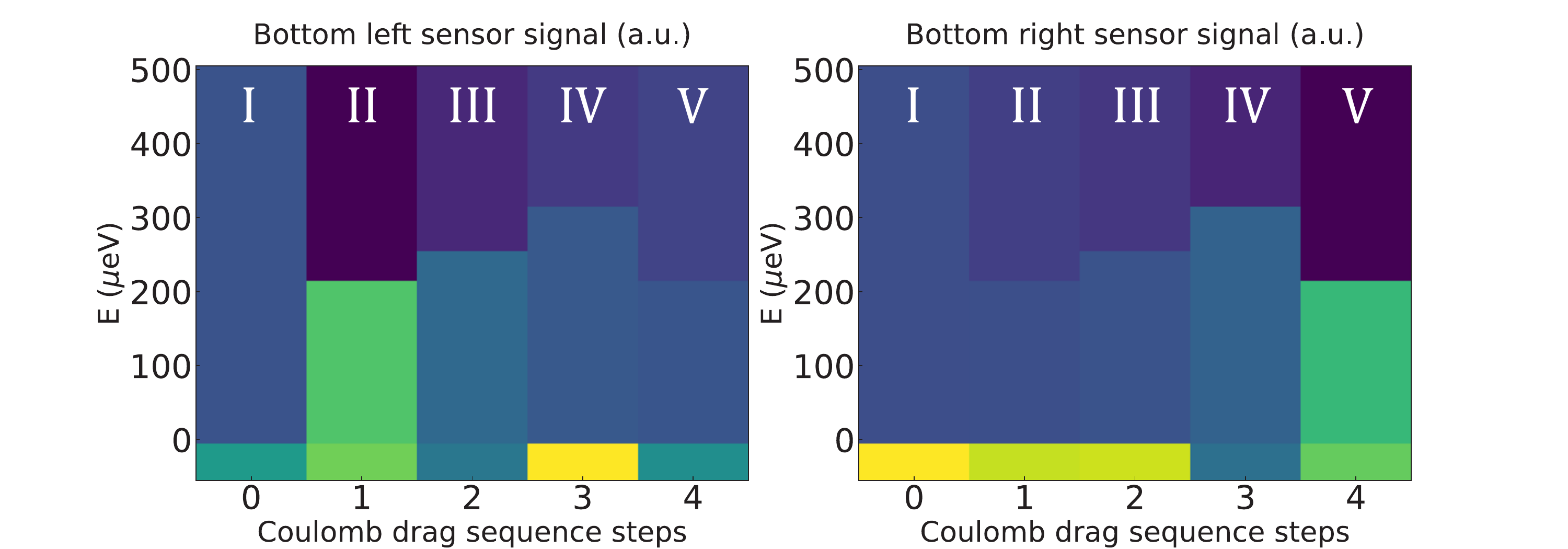}
\caption{Numerical simulation of exciton transport measurements. The simulation calculates the ground state charge configuration as a function of electrochemical potentials $\{\epsilon_i\}$. At every Coulomb drag step, the corresponding dot potential is pulsed by $\SI{670}{\micro eV}$ as specified in the main text. The simulation uses the measured $V_{ij}$ (see section IV) and assumes slightly larger intra-channel Coulomb interaction terms of $\SI{400}{\micro eV}$. Further we input $U_i = \SI{2000}{\micro eV}$ and neglect next-nearest neighbour interactions. The calculated charge state is transformed to sensor signal assuming a linear sensor response and a $\frac{1}{r^2}$ decay of interactions over distance. The simulation data is in good agreement with the measured data (Fig. \ref{CD_data_fig}a).}
\label{drag_sim_fig}
\end{figure*}

Figure \ref{drag_sim_fig} shows the simulated charge ground state variation as a function of the electrochemical potentials $\{\epsilon_i\}$. These are varied in the same way as in the experiment: The bottom (drag) channel detuning $E$ is swept from $\SI{500}{\micro eV}$ to past the Fermi energy, while the individual top channel potentials are raised and lowered by $\SI{670}{\micro eV}$, corresponding to the charge shuttling sequence specified in section IV. The charge states are converted to charge sensor signal by inputting the sensor-to-dot distances $\{r_i\}$ and assuming  $\frac{1}{r^2}$ decay of Coulomb interactions and a linear response of the sensors. The numerical simulations show good agreement with the measured data. We point out that for each dot pair, the exciton transport window is equal to the inter-channel Coulomb interaction $V_{ij}$, as highlighted in the main text. The faster vanishing response of the measured data as opposed to the numerical simulations can be explained by a decay of Coulomb interactions faster than $\frac{1}{r^2}$, as previously observed in other work \cite{Knorzer2022}.

\end{appendix}

%
\end{document}